\documentclass[fleqn,usenatbib]{mnras}
\usepackage{newtxtext,newtxmath}
\usepackage[T1]{fontenc}
\usepackage{bm}
\usepackage{ae,aecompl}
\usepackage{graphicx}
\usepackage{amsmath}
\usepackage[makeroom]{cancel}
\usepackage{color}
\usepackage{ulem}

\DeclareRobustCommand{\VAN}[3]{#2}
\let\VANthebibliography\thebibliography
\def\thebibliography{\DeclareRobustCommand{\VAN}[3]{##3}\VANthebibliography}

\title[A2029]{The Case for Hot-Mode Accretion in Abell 2029}

\author[Prasad et al.]{
Deovrat Prasad,$^{1,2}$\thanks{E-mail: deovratd@cardiff.ac.uk}
G. Mark Voit,$^{2}$
and Brian W. O'Shea $^{2,3}$
\\
$^{1}$School of Physics and Astronomy, Cardiff University, Wales, UK\\
$^{2}$Department of Physics and Astronomy, Michigan State University, MI, US\\
$^{3}$Department of Computational Mathematics, Science, and Engineering, Michigan State University, MI, US
}
\begin{document}
\label{firstpage}
\pagerange{\pageref{firstpage}--\pageref{lastpage}}
\maketitle
\begin{abstract}
Radiative cooling and AGN heating are thought to form a feedback loop that regulates the evolution of low redshift cool-core galaxy clusters. Numerical simulations suggest that formation of multiphase gas in the cluster core imposes a floor on the ratio of cooling time ($t_{\rm cool}$) to free-fall time ($t_{\rm ff}$) at $\min ( t_{\rm cool} / t_{\rm ff} ) \approx 10$. Observations of galaxy clusters show evidence for such a floor, and usually the cluster cores with $\min ( t_{\rm cool} / t_{\rm ff} ) \lesssim 30$ contain abundant multiphase gas. However, there are important outliers. One of them is Abell 2029, a massive galaxy cluster ($M_{200} \gtrsim 10^{15}$ M$_\odot$) with $\min( t_{\rm cool}/t_{\rm ff}) \sim 20$, but little apparent multiphase gas. In this paper, we present high resolution 3D hydrodynamic AMR simulations of a cluster similar to A2029 and study how it evolves over a period of 1-2 Gyr. Those simulations suggest that Abell 2029 self-regulates without producing multiphase gas because the mass of its central black hole ($\sim 5\times 10^{10} \, M_\odot$) is great enough for Bondi accretion of hot ambient gas to produce enough feedback energy to compensate for radiative cooling.
\end{abstract}

\section{Introduction}
\label{sec:intro}
In galaxy clusters, the majority of the baryons ($\sim 85\%$) are in the form of hot diffuse plasma known as the intracluster medium. 
About a third to half of the observed galaxy clusters show a cooling time ($t_{\rm cool}$) of the intracluster medium (ICM) in the core ($r<30$ kpc) smaller than the cluster's lifetime (e.g. \citealt{santos2017}). Such cool-core clusters, in the absence of any quenching mechanism, will have cold gas condensing out from the hot ICM in the order of several hundreds of M$_\odot$ yr$^{-1}$, leading to star formation at a similar rate (\citealt{fabian1994}). However, observations of cool-core clusters don't show these features, with star formation rates (SFR) in the brightest cluster galaxy's (BCG) rarely exceeding 10$\%$ of the expected value (\citealt{edge2001,peterson2003, McDonald2018}). 

Observations of cool-core galaxy clusters show the interplay between radiative cooling of the hot intracluster medium (ICM) and active galactic nuclei (AGN) feedback in the evolution of properties like star formation and gas kinematics. Both observations and numerical simulations support a scenario in which cold gas clumps condensing out of the hot ICM due to radiative cooling fuel powerful AGN outbursts that transfer enough energy back into the ICM to compensate for its radiative losses  (\citealt{birzan2004,pizz2005,rafferty2006,voit15N}). According to that scenario, cluster atmospheres become increasingly prone to producing cold clouds as the minimum ratio of cooling time ($t_{\rm cool}$) to free-fall time ($t_{\rm ff}$) declines. Those cold clouds then fuel a burst of feedback that lowers the ambient gas density and raises $t_{\rm cool}$, thereby imposing a floor at $\min ( t_{\rm cool} / t_{\rm ff} ) \sim 10$ (\citealt{mccourt12, sharma12, voit15L, voit2015, Hogan2017,babyk2018}). 

Over the past decade a generation of numerical studies have used variations of cold-mode AGN feedback to reproduce the observed features of cool-core clusters (\citealt{gaspari2012,Prasad15,li15, yang2016,  meece2017,  Prasad2018}). The success of cold-mode AGN feedback in controlling catastrophic cooling of the ICM and consequently the star formation rate (SFR) in the cool-core clusters critically depends on the episodic nature of the accreting cold gas (\citealt{gaspari2013, tremblay2016, Prasad2017, Nobels2022})  
The episodic nature of cold accretion onto the central super massive black hole (SMBH) makes the  general behavior of cold-mode feedback insensitive to accretion efficiency even though some of the features like cold-gas mass in the core differ. On the other hand, spherically symmetric and single phase hot-mode accretion is a continuous function of central entropy/density. As a consequence, the potential impact of hot-mode (Bondi) AGN feedback has a strong dependence on the AGN efficiency (\citealt{Nemmen2015}). The hot-mode feedback requires higher accretion efficiency for feedback to supply the cavity power of AGN jets in cool-core clusters like MS0735 and Centaurus (assuming $P_{\rm hot} \sim 0.1 \dot{M}_{\rm hot} c^2$; \citealt{McN2011}). 
As a result of these observations, cold-mode AGN feedback has emerged as a preferred choice for studying the evolution of cool-core clusters.

The hypothesized cold-mode AGN feedback cycle has difficulty accounting for two famous massive cool-core clusters, however. One of them is the {\it Phoenix} cluster,
which has an unusually high cooling and star formation rate ($\sim 500$ M$_\odot$ yr$^{-1}$) and $\min ( t_{\rm cool}/t_{\rm ff} ) \approx 1$ (\citealt{McDonald2019}). The other is Abell 2029 (hereafter A2029), which has very little cold gas and star formation (\citealt{martz2020}) despite having $\min ( t_{\rm cool}/t_{\rm ff} ) \approx 20$, one of the most massive central galaxies known, with $M_* \sim 2\times10^{12}$ M$_\odot$ (\citealt{Hogan2017}), and one of the most massive known central supermassive black holes, with $M_{\rm BH}\sim5\times10^{10}$ M$_\odot$ (\citealt{Brockamp2016, Dullo2017}). These massive clusters show extreme behaviour that deviates from the typical cluster population where cooling and star formation rate remains subdued (close to $10\%$ of the pure cooling flow values) with radiative cooling and AGN heating maintaining the cluster core in a dynamic thermal equilibrium (\citealt{Hogan2017, babyk2018}). Previous numerical studies have shown that the feedback cycles go through transitory ($\Delta t \sim 50-100$ Myr) pure cooling and extreme heating phases (\citealt{Prasad15,Prasad2018,prasad2020}). 
However, given that A2029 and Phoenix are both massive galaxy clusters, the question that arises is whether the overall evolution of these cool-core clusters deviates from our understanding of more typical (and generally less massive) cool-core clusters at low redshifts and if so what causes the deviations. 

In this paper, we present a numerical study of the evolution of A2029-like cool-core clusters wherein we attempt to understand the nature of its self-regulation.
The paper proceeds as follows. Section \ref{sec:methods} introduces the initial cluster setup and physics modules in the simulation. Section \ref{sec:results} presents the main results of the simulation.
Section \ref{sec:disc} discusses whether our results can be considered realistic, explores their dependence on different assumptions made in the simulations and compares them to observations of A2029. Section \ref{sec:conc} summarizes the paper's main findings.

\section{Numerical Methods}
\label{sec:methods}
We modified \textsc{Enzo}, an adaptive mesh refinement (AMR) code \citep{Bryan2014,Enzo_2019}, to simulate AGN and stellar feedback in idealized galaxy cluster environments similar to A2029. The mass of the simulated halos is $M_{200} \sim 10^{15} M_{\odot}$, where $M_{200}$ is the mass contained within the radius $r_{200}$ encompassing a mean mass density 200 times the cosmological critical density.
We solve the standard hydrodynamic equations in Cartesian coordinates, including radiative cooling, gravity, star formation, stellar feedback, and AGN feedback as described in \citet{prasad2020}.
These simulations employ the piecewise parabolic method (PPM) with a Harten-Lax-Van Leer-Contact (HLLC) Riemann solver. 

\subsection{Grids}
\label{sec:bc}
The simulation domain is a $(4\ {\rm Mpc})^3$ box with a $64^3$ root grid and up to 10 levels of refinement. The central (128 kpc)$^3$ region enforces static regions of grid refinement ranging from level 6 to level 9, with the refinement level increasing toward the center.  In the central $(2 \, {\rm kpc})^3$, the mesh is fixed to be at the highest level of refinement. 
This design ensures that the CGM is highly resolved at all times with a minimum cell size $\Delta x \approx 61 \, {\rm pc}$ and a maximum cell size $\Delta x < 1$ kpc within $r<60$ kpc.

\subsection{Cluster Parameters}
\label{sec:gpt}
The gravitational potentials in our simulations have three components, and they do not change with time.  We use an NFW form (\citealt{nav1997}) for the dark-matter halo, with mass density $\rho_{\rm DM} \propto r^{-1} (1 + r/r_{\rm s})^{-2}$, where $r_{\rm s}$ is the NFW scale radius and $c_{200} = r_{200} / r_{\rm s} = 5.0$ is a halo concentration parameter.  For the central galaxy, we use a Hernquist profile (\citealt{hern1990}), with a stellar mass density $\rho_* \propto r^{-1} (1 + r/r_{\rm H})^{-3}$, where $r_{\rm H}$ is the Hernquist scale radius ($r_H = 20$ kpc for our simulations) and mass of central galaxy, $M_{\rm BCG}\approx 2\times10^{12}$ M$_\odot$ (\citealt{Hogan2017}). The potential of the central supermassive black hole of mass $M_{\rm BH}$ follows a Paczynski-Witta form (\citealt{pacwita1980}) with $M_{\rm BH}=5\times 10^{10}$ M$_\odot$ (\citealt{Dullo2017}). 

\subsection{Star formation and Stellar Feedback}
\label{sec:Afeed}
The simulations use the \citet{cenos92} prescription for star formation in Enzo as described in \citet{prasad2020}. Feedback from stars formed during the simulation follows the prescription from \citet{Bryan2014}. After a star particle of mass $m_*$ forms, a fraction $f_{m,*}$ of its mass is added back to the cell, along with thermal energy $E_{\rm SN} = f_{\rm SN} m_* c^2$. Our simulations adopt the parameter values $f_{\rm SN}=1\times10^{-5}$ and $f_{m,*}= 0.25$. The returned gas formally has a metallicity $f_{Z,*}=0.02$, which can be used as a passive tracer but is not included in our radiative cooling calculations which uses the \citet{dopita1993} cooling table assuming a fixed metallicity of $Z=0.3 Z_\odot$.  This process starts immediately after the formation of the star particle and decays exponentially, with a time constant of 1~Myr. 

\subsection{AGN Feedback}
\label{sec:agn}
AGN feedback is introduced into the simulation using a feedback zone attached to the AGN particle (\citealt{meece2017}), which is always located at the geometric center of the halo. We drive AGN feedback in the form of a bipolar outflow by putting operator-split source terms for mass, momentum and energy into the fluid equations as follows:
\begin{eqnarray}
\label{eq:fl_eq}
\frac {\partial \rho} { \partial t} + \nabla \cdot (\rho {\bf u}) & \, = \, & S_{\rho} \\
\frac {\partial (\rho {\bf v})} { \partial t} + \nabla \cdot (\rho {\bf v} \otimes {\bf v}) & \, = \, & -\nabla P - \rho\nabla \Phi + S_{\bf p} \\
\frac {\partial (\rho e)} {\partial t} + \nabla \cdot [(\rho e +P){\bf v}] &  \, = \, & \rho{\bf v \cdot g} - n_e n_i \Lambda(T,Z) + S_{e} 
\end{eqnarray}
where $S_{\rho}$, $S_{\bf p}$ and $S_{e}$ are the density, momentum and energy source terms, respectively.  The specific energy $e$ includes kinetic energy, and the corresponding equation of state is 
\begin{equation}
    P = (\gamma -1) \rho (e - v^2/2)
\label{eq:eos}
\end{equation}
where, $\gamma = 5/3$ for the ideal gas.
\subsubsection{Mode of Accretion}
\label{sec:acc}
The accretion rate $\dot{M}_{\rm acc}$ onto the central supermassive black hole is calculated by assuming that all the cold gas ($T<10^5$ K) within $r< 0.5$ kpc accretes onto the central black hole on a 1~Myr time scale.  
This mass accretion rate fuels AGN feedback at an energy output rate given by:
\begin{equation}
\dot{E}_{\rm AGN} = \epsilon \dot{M}_{\rm acc} c^2
\; \; ,
\label{eq:power}
\end{equation}
where $c$ is the speed of light and $\epsilon$ is the feedback efficiency parameter. A cold gas mass equal to  $\dot{M}_{\rm acc} \Delta t$ is removed from the spherical accretion zone ($r < 0.5$ kpc) by subtraction of gas mass from each cell in that zone that has a temperature below $10^5$~K, and the amount of gas mass subtracted from the cell is proportional to its total gas mass. 

The hot-mode or Bondi accretion rate is modelled assuming the SMBH mass M$_{\rm BH}\approx 5\times10^{10}$ M$_\odot$ (\citealt{Dullo2017}) with hot mode accretion rate given by
\begin{equation}
\label{eq:hot}
\dot{M}_{\rm hot} = \frac{4\pi G^2 M_{\rm BH}^2 \rho_{\rm mean}}{c_s^3}
\end{equation}
where $\rho_{\rm mean}$ is the mean density within the Bondi radius, $R_{\rm Bondi}$. 
For the run with hot+cold mode accretion, the hot-mode accretion rate is summed up with the cold mode accretion rate to give a total rate, i.e., $\dot{M}_{\rm acc} = \dot{M}_{\rm cold} + \dot{M}_{\rm hot}$.

\subsubsection{AGN Feedback Efficiency}
\label{sec:efficiency}
An AGN efficiency factor accounts for the gas within the accretion zone being able to reach the SMBH sphere of influence and the mass-energy conversion for the accreting gas. For our simulations in \citet{prasad2020} we used a relatively low AGN efficiency factor, $\epsilon=10^{-4}$, guided by earlier cold-mode AGN feedback simulations in cool-core clusters (\citealt{Prasad15, Prasad2018}). However, smaller AGN efficiency results in stronger momentum flux per unit jet power in our simulations because of the way AGN feedback has been implemented (\citealt{Prasad2022}). 

A major consequence of this choice is that AGN feedback leads to large scale atmospheric circulation, which in turn leads to elevated entropy in the cluster cores. For the current suite of our cold-mode AGN feedback simulations in A2029-like galaxy clusters, we have chosen the AGN efficiency, $\epsilon$, to be $3\times10^{-3}$ and $10^{-2}$. The choice of $\epsilon=0.003$ is guided by observations of the $M_{\rm BH}-M_{\rm halo}$ relation (\citealt{Booth2009, Megan2022}). These choices also allow us to compare with other numerical studies of evolution of cool-core clusters at low redshift. For the `Hot+Cold' mode feedback, the jet power is calculated according to Equation \ref{eq:power} with accretion efficiency $\epsilon=0.01$ for the total accretion rate. As such the pure hot-mode (Bondi) jet power is $P_{\rm hot} = 0.01 \dot{M}_{\rm hot} c^2$.  

\subsubsection{Feedback Energy Deposition}
\label{sec:part}
The AGN output energy is partitioned into kinetic and thermal parts and introduced using the source terms described above. The source regions are cylinders of radius 0.5~kpc extending along the jet axis from $r = 0.5$~kpc to $r = 1$~kpc in each direction. Each jet therefore subtends 0.93~radian at $r = 1$ kpc.

In each cell of volume $\Delta V_{\rm cell}$ within the source region of volume $V_{\rm jet}$, the density source term is $S_\rho = \Delta m_{\rm cell} / \Delta V_{\rm cell}$, where $\Delta m_{\rm cell} = (\Delta V_{\rm cell} / V_{\rm jet}) \dot{M}_{\rm acc} \Delta t$ and $\Delta t$ is the time step.  
Within the source region, the corresponding energy source term in each cell is $S_e = (\Delta e_{\rm cell} + \Delta {\rm KE}_{\rm cell})/V_{\rm cell}$, where 
\begin{equation}
   \Delta e_{\rm cell} = (1 - f_{\rm kin}) \dot{E}_{\rm AGN} \Delta t \cdot \frac{\Delta V_{\rm cell}}{V_{\rm jet}} 
\end{equation}
\begin{equation}
   \Delta {\rm KE}_{\rm cell} = f_{\rm kin} \dot{E}_{\rm AGN} \Delta t \cdot \frac{\Delta V_{\rm cell}}{V_{\rm jet}}
   \; \; .
\end{equation}
For all our runs the ratio of kinetic to total AGN output energy is fixed at $f_{\rm kin} = 0.9$.

The momentum source term in each cell is $S_{\bf p}= \Delta {\bf p}_{\rm cell}/ \Delta V_{\rm cell}$, where
\begin{equation}
    \Delta {\bf p}_{\rm cell} = \Delta m_{\rm cell} 
        \sqrt{ \frac{2\Delta {\rm KE}}{{\Delta m_{\rm cell}}} } \hat{\bf n}
\end{equation}
and $\hat{\bf n}$ is a unit vector pointing away from the origin along the jet injection axis. 

\subsection{ICM initialisation}
\label{sec:icm}
The simulations have been initialised with hydrostatic, single-phase galactic atmospheres having profiles of density, pressure, and specific entropy consistent with observations of A2029 by \citet{Hogan2017}.  The red-dashed lines in Figure~\ref{fig:evol_fid} shows the initial atmospheric properties for the simulation. The initial entropy profile is modelled using the form $K(r)=K_0 + K_{100} (r/100 \, {\rm kpc})^\alpha$ (\citealt{cavagnolo09}) with $K_0=20.0$, $K_{100}=110$ and $\alpha =1.1$ for the halo. Throughout the simulation, atmospheric gas is allowed to cool to $10^3$ K using tabulated \cite{SD1993} cooling functions assuming one-third Solar metallicity. 

\section{Results}
\label{sec:results}
The expectation was that Abell 2029, like a typical galaxy cluster, would go through different stages of evolution with a cooling phase and a higher min($t_{\rm cool}/t_{\rm ff}$) phase as the AGN jets become active. Our simulation results, however, have led us to a different interpretation of the feedback cycle in A2029.
In this section, we describe the key results from our simulations. Table \ref{tab:runs} lists the three simulation runs for the A2029 cluster. This is followed by results where we show the spherically-averaged radial profiles for electron density, entropy, cooling time ($t_{\rm cool}$) and cooling time to free-fall time ratios ($t_{\rm cool}/t_{\rm ff}$) for all the three runs and do a comparative study. We show the evolution of gas mass in different temperature ranges along with jet power. We also explore the different nature of energy and mass fluxes within the central $r<30$ kpc of the cluster core for all the three runs. 
\begin{table}
\caption{List of runs}
\resizebox{0.480 \textwidth}{!}{
\begin{tabular}{c c c c c}
\hline
Run   & cold-Mode  & hot-mode & accretion & run time \\
      & accretion & accretion & efficiency ($\epsilon$) & (Gyr)\\
\hline
A &  yes & No & 0.003 & 1.5 \\
B &  yes & No & 0.01 & 0.5 \\
C &  yes & Yes & 0.01 & 1.0 \\
\hline
\end{tabular}}
\label{tab:runs}
\\
\end{table}

\subsection{Cold-Mode Feedback, $\epsilon=0.003$}
\label{sec:fid_run}

\begin{figure*}
 \includegraphics[width=1.0\textwidth]{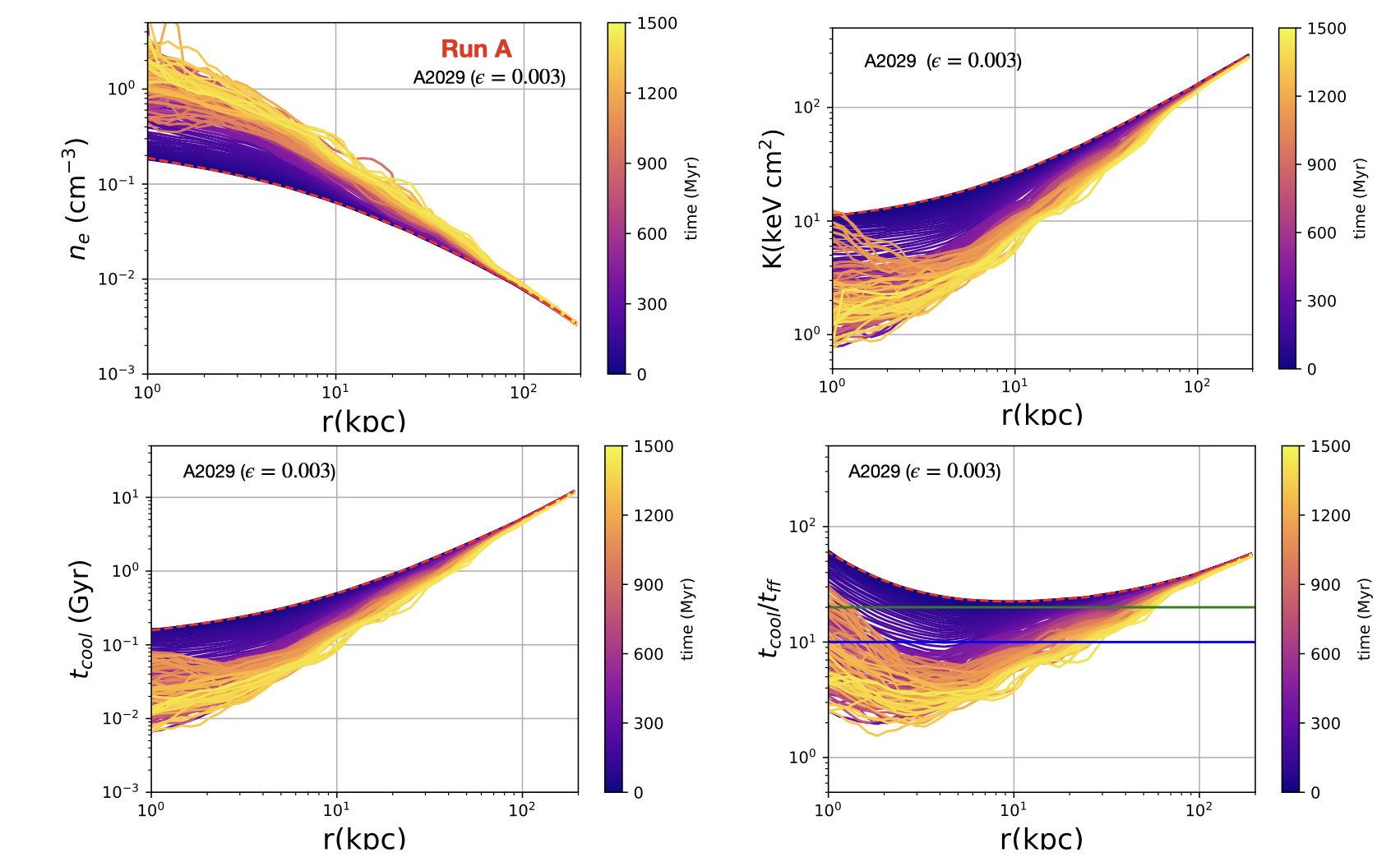}
 \caption{Mass-weighted spherically averaged radial profiles of electron density (top left), entropy (top right), cooling time (bottom left) and $t_{\rm cool}/t_{\rm ff}$ (bottom right) with time ($t=0-1.5$ Gyr) for the cold-mode feedback Run A with feedback efficiency $\epsilon=0.003$. Each line shows the respective quantities with a 10 Myr cadence. Red-dashed lines show the initial state. The radial profiles show that the core of the simulated cluster settles down in a persistent state with low $t_{\rm cool}/t_{\rm ff}$ making the ICM susceptible to thermal instability.}
 \label{fig:evol_fid}
\end{figure*}
We experimented with the accretion efficiency ($\epsilon$) and mode of AGN fuelling (see Table \ref{tab:runs}). For Run A, we chose the accretion efficiency $\epsilon = 0.003$ for cold-mode AGN fuelling/triggering. This choice of the accretion efficiency allows us to better compare with other similar numerical studies of cool-core clusters. 
Figure \ref{fig:evol_fid} shows the spherically averaged radial electron density (top left panel), entropy (top right panel), cooling time ($t_{\rm cool}$; bottom left panel) and cooling time to free-fall time ($t_{\rm cool}/t_{\rm ff}$; bottom right panel) of the baryons over $t=1.5$ Gyr for Run A. The red-dashed lines show the initial radial profile for the respective quantity. As the cluster evolves the central density (top left panel) keeps  rising within the central $r<30$ kpc, with min($t_{\rm cool}/t_{\rm ff}$) falling below 10 by 0.3 Gyr (bottom right panel). This leads to a monotonic drop in core entropy ($K < 10$ keV~cm$^2$ within $r< 10$ kpc). 
AGN feedback is unable to evacuate the high density gas from the deep potential well, with the cluster core staying with min$(t_{\rm cool}/t_{\rm ff})\lesssim 5$ for the remainder of the simulation. The monotonic rise in the central ($r<30$ kpc) gas density leads to frequent jet outbursts.  

\begin{figure*}
\centering
 \includegraphics[width=0.45\textwidth]{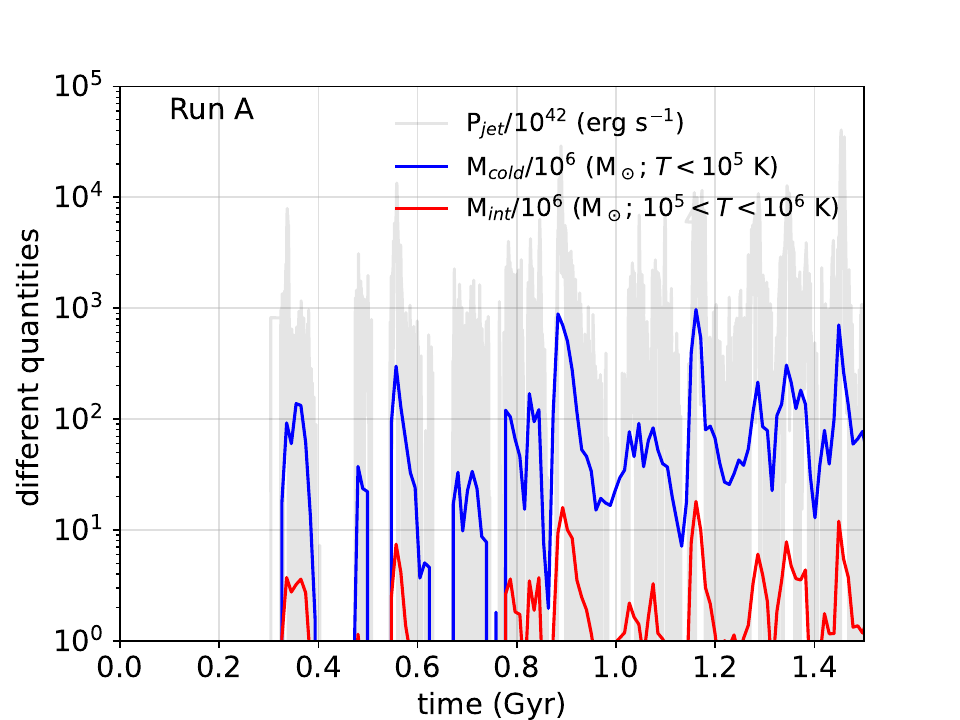}
 \includegraphics[width=0.45\textwidth]{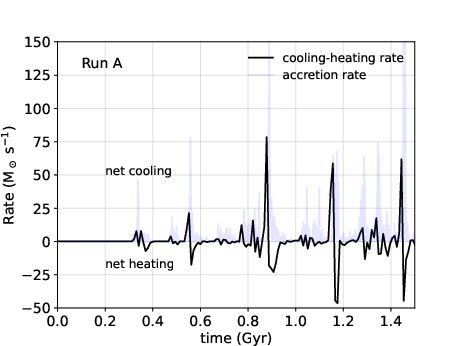} 
 \caption{  {\it Left panel} : Jet power ($P_{\rm jet}$) and cold gas mass with time for Run A. The background grey lines show the instantaneous injected jet power while blue line shows the evolution of the cold gas mass ($T < 10^5$ K) and red line shows the evolution of the intermediate temperature ($10^5<T<10^6$ K) gas with time for Run A. For more than 1.5 Gyr the cold gas mass stays within the upper limits of \citet{Hogan2017} for A2029 despite the cluster core ($r<20$ kpc) persisting with min$(t_{\rm cool}/t_{\rm ff}) < 10$ and a central density exceeding observations by an order of magnitude. 
 {\it Right panel} : Net Cooling rate (solid black line) of gas below $10^6$ K within central $r<50$ kpc and accretion rate (background purple line) on the SMBH ($r<0.5$ kpc) with time for Run A. 
}
 \label{fig:fid_gas_mass}
\end{figure*}
The left panel in Figure \ref{fig:fid_gas_mass} shows the 
instantaneous jet power with time (background grey lines). It shows that the AGN activity turns on at $t \approx 300$ Myr. From Figure \ref{fig:evol_fid} (bottom left panel) we know that  the min($t_{\rm cool}/t_{\rm ff}$) falls below 10 by $t\sim300$ Myr with the cooling time of gas within $r<10$ kpc falling to $t \lesssim 100$ Myr. Consequently, the cooling rate rises and the accretion rate shoots up, resulting in the AGN jet outburst power peaking at P$_{\rm jet}\sim10^{45}$ erg s$^{-1}$. 
The total cold gas (T$<10^5$ K) mass in the domain (blue line, left panel) reaches about $10^8$ M$_\odot$ at about $t\sim 300$ Myr. The buildup of cold gas coincides with the AGN activity in the simulation. The AGN outburst heats the ICM, lowering the accretion rate, which in turn lowers the jet power. The first jet cycle lasts for about 100 Myr. This pattern of cooling-heating cycle (right panel) is repeated over the evolution of the cluster.

The net cooling-heating rate (solid black line) and the accretion rate (background purple line) for Run A  is shown in the  right panel of Figure \ref{fig:fid_gas_mass}. This plot shows that at about $t\sim 300$ Myr the cooling rate picks up leading to a sharp rise in the accretion rate onto the SMBH. This leads to the first AGN outburst, which results in net heating of the ICM and a decrease in the accretion rate. This pattern is repeated over the entire course of the simulation.  
The cold gas mass does not accumulate in the cluster core (middle panel, Figure \ref{fig:fid_gas_mass}) despite conditions being conducive for multiphase condensation (low min($t_{\rm cool}/t_{\rm ff}$) and high central density). This is likely because the cold gas clumps condensing out of the hot ICM fall within the accretion zone rapidly since the free-fall time, $t_{\rm ff}$, is small ($\lesssim 10$ Myr) within $r<3$ kpc. This rapid infall of the condensing cold gas clumps leads to frequent sharp spikes in accretion rate as seen in the cooling/heating rate plot resulting in frequent high jet power AGN outbursts. AGN feedback seems to be `preventative' rather than `ejective' as the cluster core stays in this dynamic heating-cooling equilibrium over $t\sim 1.5$ Gyr of the cluster evolution. 

\subsection{Higher efficiency cold-mode feedback, $\epsilon=0.01$ }
\label{sec:hiff}
\begin{figure*}
 \includegraphics[width=1.0\textwidth]{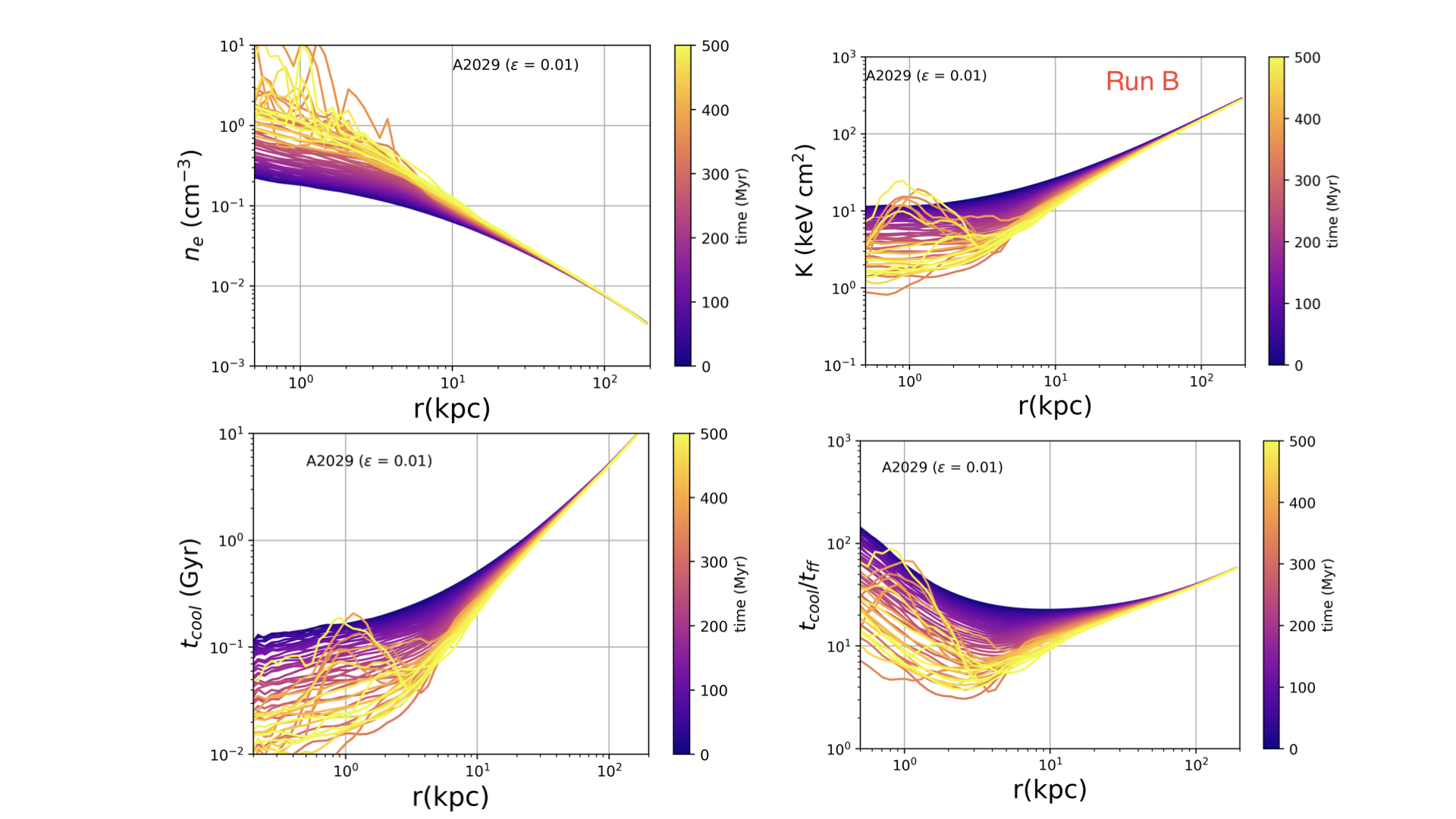}
 \caption{ Mass-weighted and angle-averaged radial profile electron density (top left, entropy (top right), cooling time ($t_{\rm cool}$; bottom left) and $t_{\rm cool}/t_{\rm ff}$ (bottom right) with time ($t=0-0.5$ Gyr) for cold-mode feedback Run B with feedback efficiency $\epsilon=0.01$. These radial profiles show that the cluster core ($r<30$ kpc) settles down in a state of low $t_{\rm cool}/t_{\rm ff}$ within the core with a greater than observed central gas density. Increasing the accretion efficiency does not stop a monotonic rise of the baryon density within $r<10$ kpc. The cluster core evolution is similar to Run A ($\epsilon=0.003$).}
 \label{fig:evol_heff}
\end{figure*}   
Figure \ref{fig:evol_heff} shows the evolution 
of the electron density (top left panel), entropy (top right panel), cooling time (bottom left panel), and $t_{\rm cool}/t_{\rm ff}$ (bottom right panel) for the cold-mode feedback run with high accretion efficiency, $\epsilon=0.01$. These plots show results which are similar to Run A: a monotonic rise in the gas density within central $r<30$ kpc. Entropy, cooling time ($t_{\rm cool}$), and $t_{\rm cool}/t_{\rm ff}$ show similar behaviour to Run A even though the initial jet outburst does raise the entropy and $t_{\rm cool}$ briefly ($\Delta t \sim 50$ Myr) in the central $r<5$ kpc. 

Increasing $\epsilon$ to $0.01$ causes the injection jet velocity, $v_{\rm jet}$ to increase from $\sim 1.6\times10^4$ km s$^{-1}$ (for $\epsilon=0.003$) to $\sim 3\times10^4$ km s$^{-1}$ ($v_{\rm jet} =  \sqrt{\epsilon c^2}$ as $\dot{M}_{\rm jet}=\dot{M}_{\rm acc}$ for our simulations). This brings down the jet momentum flux per unit power of the jet, leading to better coupling with the ICM, as shown in \citet{Prasad2022}. However, given the massive central galaxy residing in one of the most massive cluster halos, the central potential is too deep for the AGN jets to cause massive evacuation of the gas buildup during the initial cooling phase ($t=0.3$ Gyr) in the core ($r<30$ kpc). Consequently, increasing the accretion efficiency does not prevent the simulation's central gas density from exceeding observations of A2029. As the simulation progresses further, the gas density continues to show a monotonic rise similar to Run A. We therefore terminate the Run B at $t=0.5$ Gyr. 

\subsection{The Hot+Cold mode accretion}
\label{sec:bac}
\begin{figure*}
 \includegraphics[width=0.95\textwidth]{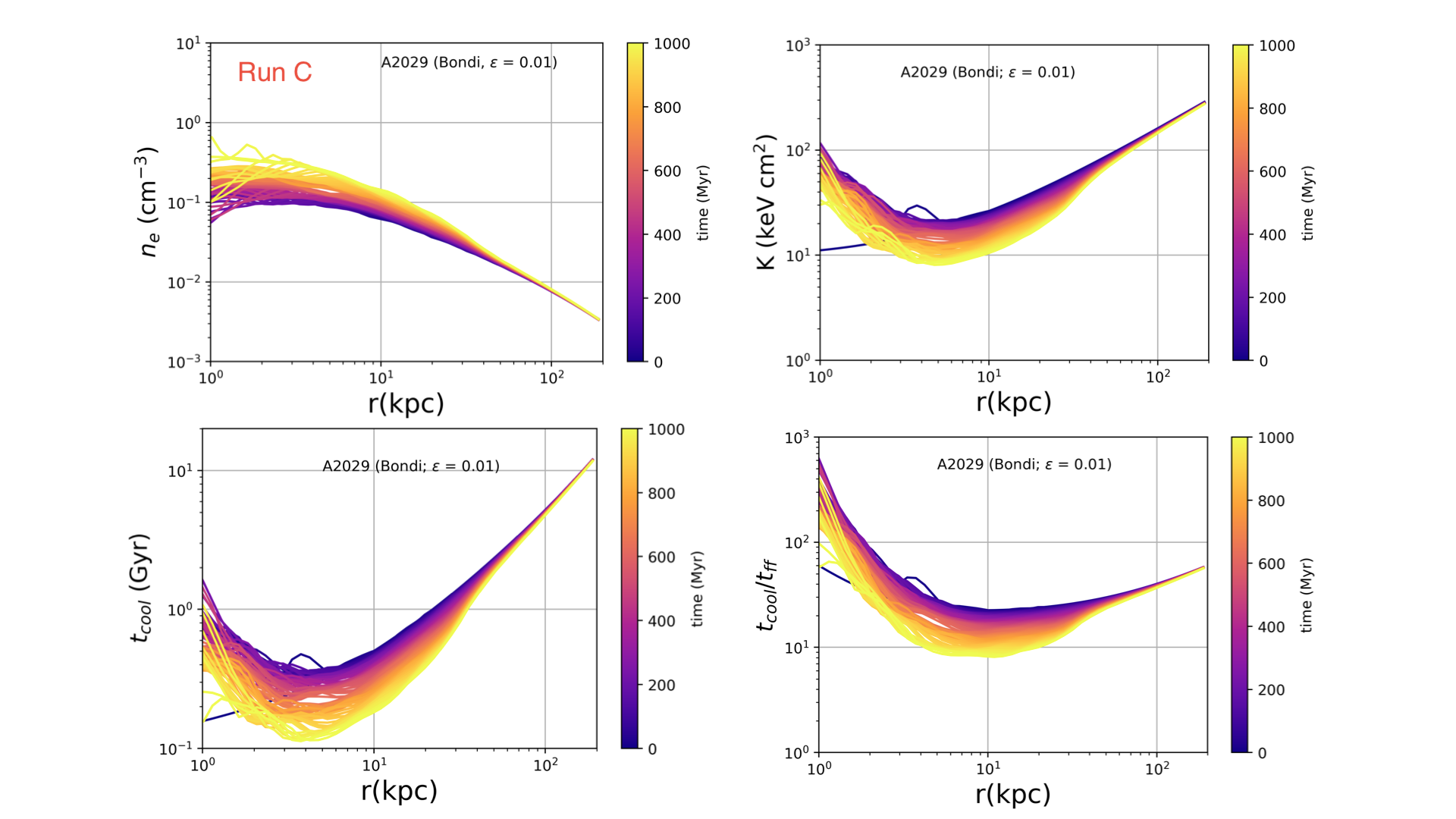}
 \caption{ Mass weighted and angle-averaged radial profile for electron density (top left), entropy (top right), $t_{\rm cool}$ (bottom left) and $t_{\rm cool}/t_{\rm ff}$ (bottom right) with time ($t=0-1.0$ Gyr) with a 10 Myr cadence for `hot+cold' mode AGN feedback run with accretion efficiency $\epsilon=0.01$. 
 Plots show that the cluster core density ($r<30$ kpc) maintains the initial density profile for the initial $t\sim 300$ Myr unlike the fiducial run. This is because the hot-mode AGN feedback causes the AGN activity to switch on at the start of the simulation preventing the gas from accumulating in the core.  }
 \label{fig:evol_prof_hot}
\end{figure*}
In this section we explore the role of hot-mode AGN feedback in addition to cold-mode feedback (Run C). Hot-mode AGN feedback can be significant in clusters with a very massive SMBH as in A2029 ($M_{\rm SMBH}\sim 5\times10^{10}$ M$_\odot$) because the Bondi accretion rate is $\propto M_{\rm BH}^2$. Hot-mode accretion and AGN feedback is included in the simulations as described in Section \ref{sec:acc}.
Enabling the hot-mode accretion and AGN feedback causes the AGN activity to switch on at the start of the simulation. 

Figure \ref{fig:evol_prof_hot} shows the evolution of electron density (top left panel), entropy (top right panel), cooling time (bottom left panel), and $t_{\rm cool}/t_{\rm ff}$ (bottom right panel) for $t=0-1$ Gyr for the Hot+Cold mode AGN feedback run. Plots show that enabling the AGN feedback at the start of the simulation (i.e., by enabling hot mode accretion) prevents the buildup of gas density that happens during the initial pure cooling phase ($t\sim 300$ Myr) in Run A and Run B. All the quantities show relatively minor changes over $t=1$ Gyr compared to the runs with only cold-mode feedback, with electron density staying close to the observed values for A2029. The steady hot-mode feedback pushes the entropy, cooling time, and $t_{\rm cool}/t_{\rm ff}$ to higher values in the center ($r<3$ kpc) over $t\sim 1$ Gyr. 
\begin{figure}
 \includegraphics[width=0.48\textwidth]{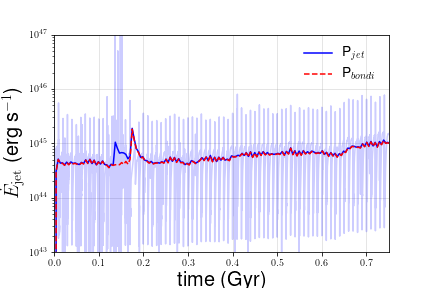}
 \caption{Total jet power (P$_{\rm jet}$; solid blue line) averaged over $\delta t=10$ Myr with time for the `Hot+Cold' mode run. A dashed red line shows the hot-mode jet power, $P_{\rm Bondi}$ ($= \epsilon \dot{M}_{\rm Bondi c^2}$), with time for accretion efficiency $\epsilon=0.01$, same as the cold mode feedback. The background cyan line represents the instantaneous total injected P$_{\rm jet}$ with time. Plots show that hot-mode AGN activity is enabled from the beginning of the simulation and dominates over the cold-mode feedback for most of the simulation run time.}
 \label{fig:pjet_hot}
\end{figure}
Figure \ref{fig:pjet_hot} shows the total jet power (solid blue line) smoothed over a small window of $\Delta t=10$ Myr for Run C. The dashed red line shows the average Bondi (hot-mode) jet power over the same time while the background cyan line shows instantaneous total jet power for the hot+cold mode run. The hot-mode accretion enables AGN activity from the start with $P_{\rm jet}\sim$ few times $10^{44}$ erg s$^{-1}$. The steady AGN activity keeps the cooling gas from accumulating in the core. For most of the simulation time ($\Delta t \sim 0.75$ Gyr) the jet power ($P_{\rm jet}$) is dominated by hot-mode accretion. $P_{\rm jet}$ also shows a slight rising trend as the central density gradually rises over time (top left panel, Figure \ref{fig:evol_prof_hot}). 

\section{Discussions}
\label{sec:disc}
The cold-mode AGN feedback prescription is a useful framework for understanding the evolution of galaxies, groups and clusters at low redshifts. However, recent observations of the {\it Phoenix} cluster by \citet{McDonald2019} showing its nearly cooling flow-like behaviour as opposed to the expected quenched star formation state shines a light on the limitations of cold-mode AGN feedback. A2029 is another such massive galaxy cluster at the other extreme of the cool-core cluster population, with little to no star formation and cold gas mass despite having a massive BCG and SMBH (\citealt{Hogan2017}).  

Several numerical studies have been able to reproduce many of the the observed features of cool-core clusters using cold-mode AGN feedback (\citealt{gaspari2012, li15, Prasad15, Prasad2018,meece2017,yang2016, Nobels2022}). The stochastic nature of cold gas accretion onto the SMBH makes the feedback episodic with short ($\sim 30-50$ Myr) bursts of jet power exceeding $10^{46}$ ergs$^{-1}$. Such powerful jet activity is able to eject and heat the cooling ICM and limit the overall cooling rate. However, a typical galaxy cluster has a central SMBH mass which is an order of magnitude smaller than the one in A2029. Given that the hot-mode jet power scales as $M_{\rm BH}^2$ (Equation \ref{eq:hot}), for a typical galaxy cluster the cold-mode AGN feedback likely remains dominant at all times.  
However, for a feedback efficiency of $\epsilon \sim 0.01$, the hot-mode (`Bondi') jet power ($P_{\rm hot} \sim \epsilon $M$_{\rm hot} c^2$) for the massive SMBH in A2029 is around $10^{45}$ erg s$^{-1}$, comparable to the typical jet power achieved for cold-mode AGN feedback (\citealt{Prasad15, Prasad2018}) and also to the cluster's radiative loss rate.

Our cold mode AGN simulations of A2029 fail to prevent the central gas density from exceeding the observed levels in the core ($r<30$ kpc) even though they do prevent cooling from becoming catastrophic. Raising the accretion efficiency does not keep the central core density within the observational limits. However, enabling the hot-mode AGN feedback allows the jets to be active from the start of the simulations and to prevent the central gas density from rising too high during the initial cooling phase. The cluster core settles into an equilibrium state in which the cold gas remains within observational limits even when min($t_{\rm cool}/t_{\rm ff})\sim10$. 
Our hot-mode AGN feedback algorithm assumes that the accretion onto the SMBH happens at `Bondi' rate (Equation \ref{eq:hot}). However, that might not happen around real AGN. For example, multiple observations of galactic centers such as Sag A* (\citealt{bag2003}), M87 (\citealt{Russell2015}), NGC 33115 (\citealt{Wong2014}), and M84 (\citealt{Bambic2023}) indicate that the current accretion rate on to the SMBH is much lower than the inferred `Bondi' accretion rate. Because of this uncertainty, the results of our simulations do not conclusively demonstrate that hot-mode accretion alone provides enough power to compensate for radiative cooling of the core gas in Abell 2029, only that it is a plausible source for the accretion power, given the feedback efficiency factor $\epsilon=0.01$ that we have assumed. It is also possible for hot accretion at somewhat less than the Bondi rate to produce enough power if the efficiency factor $\epsilon$ is somewhat greater than 0.01.

In \citet{Prasad2020P}, we had argued that the unusual properties of the {\it Phoenix} cluster might be a transient phase in the lifetime of a massive cool-core cluster. One of the aims of this study was to let the A2029-type cluster (a massive cool-core cluster with M$_{200} \gtrsim 10^{15}$ M$_\odot$) go through its evolution cycle over 1-2 Gyr interspersed with cooling and AGN heating phases. While the cluster core in our simulation does go through the feedback cycle, the local environment (i.e., the massive central galaxy and SMBH) affects the overall nature of this cycle. The massive BCG (M$_{\rm BCG}=10^{12}$ M$_\odot$) in a $10^{15}$ M$_\odot$ halo (\citealt{Dullo2017}) makes the central potential deep and episodic cold-mode AGN feedback finds it difficult to expel the gas from the core ($r<30$ kpc). As such, the central baryon density shows a secular rise over time with SMBH feedback responding to it with frequent jet outbursts.    

\section{Conclusions}
\label{sec:conc}
We have simulated a cluster setup similar to Abell 2029 with cold mode and hot+cold mode AGN feedback. Three different simulations, Runs A,B, and C, were performed with the same galaxy cluster setup but with different AGN feedback parameters. Run C has hot+cold (Bondi) mode accretion while the other two runs have just cold mode accretion feeding the AGN feedback. These simulations have allowed us to distinguish between the role of different modes of accretion and AGN feedback in simulations of massive galaxy clusters. The main results are as follows:
\begin{itemize}
\item Cold-mode feedback is able to attain a heating-cooling balance in the cluster core, resulting in low cold gas mass over nearly 2 Gyr in Run A with feedback efficiency $\epsilon = 0.003$. However, it fails to maintain the central ($r<20$ kpc) gas density at the observed level. The central gas density rises steeply leading to very short cooling times, low $t_{\rm cool}/t_{\rm ff}$, and frequent powerful jet outbursts.
\item Raising the accretion efficiency to $0.01$ in Run B does not prevent the central gas density from rising steeply. This is because during the initial 300 Myr before cold mode feedback switches on, the central density rises as gas settles to the center on account of the central galaxy's potential well being very deep (which is deepened further by the particularly massive SMBH). In our simulations of A2029 the cold mode feedback is unable to prevent the buildup of gas in the core.  
\item Enabling hot-mode accretion in addition to cold mode accretion enables feedback to switch on immediately in Run C. It prevents the central density from rising steeply before cooling picks up and cold mode feedback switches on.
\end{itemize}

\section*{acknowledgments}
DP is supported by the Royal Society through grant RF-ERE-210263 (Freeke van de Voort as PI). DP also acknowledges the support from {\it Chandra} theory grant no. TM8-19006X (G. M. Voit as PI) and NSF grant no. AST-1517908 (B.W.O'Shea as PI). DP would also like to acknowledge useful discussions with Freeke van de Voort (CU). GMV and BWO are partially supported by grant AAG-2106575 from the NSF. BWO acknowledges further support from NSF grant AST-1908109 and NASA ATP grants NNX15AP39G and 80NSSC18K1105.  The simulations were carried out in part using ACCESS resources using allocation TG-AST190022 (PI: Deovrat Prasad). This work was supported in part through computational resources and services provided by the Institute for Cyber-Enabled Research at Michigan State University.

\section*{Data Availability}
The data underlying this article will be shared on reasonable request to the corresponding author.
\bibliographystyle{mnras}
\bibliography{reference}
\label{lastpage}
\end{document}